\documentclass[notitlepage, 
noruninaddress,
bibnotes,
prb,5pt, 
twocolumn,
showpacs,preprintnumbers,superscriptaddress,
amsmath,amssymb,floatfix]{revtex4-1}
\usepackage{amsmath}
\usepackage{graphicx}
\usepackage{dsfont}
\usepackage{dcolumn}
\usepackage{bm}
\usepackage{color}
\usepackage{xcolor}
\usepackage{url}
\usepackage{tabularx}
\usepackage{array}
\usepackage{booktabs}
\usepackage{comment}
\usepackage{hyperref}
\usepackage{footnote}
\usepackage{lipsum}
\usepackage[normalem]{ulem}

\definecolor{colorFNR}{rgb}{0.1, 0.7, 0.1}

\newcounter{firstbib}

\usepackage{titlesec}
\newcommand*{\justifyheading}{\raggedright}
\titleformat{\section}
  {\normalfont\bfseries\justifyheading}{\thesection}{1em}{}  
\usepackage[singlelinecheck=false, justification=RaggedRight, format=plain]{caption}
\DeclareCaptionLabelSeparator{bar}{ $\boldsymbol{\mid}$ }
\newcommand{\thesupplementbibliography}{\thebibliography}
\usepackage{etoolbox}
\makeatletter
\apptocmd{\thesupplementbibliography}{\global\c@NAT@ctr 30\relax}{}{}
\makeatother

\begin{document}

\title{Embedded skyrmion bags in thin films of chiral magnets}

\author{Luyan Yang}
\affiliation{Beijing Key Laboratory of Microstructure and Property of Advanced Materials, College of Materials Science and Engineering, Beijing University of Technology, Beijing 100124, China}
\affiliation{Ernst Ruska-Centre for Microscopy and Spectroscopy with Electrons and Peter Gr\"unberg Institute, Forschungszentrum J\"ulich, 52425 J\"ulich, Germany}

\author{Andrii S. Savchenko}
\email{a.savchenko@fz-juelich.de}
\affiliation{Peter Gr\"unberg Institute and Institute for Advanced Simulation, Forschungszentrum J\"ulich and JARA, 52425 J\"ulich, Germany}

\author{Fengshan~Zheng}
\email{zhengfs@scut.edu.cn}

\affiliation{Spin-X Institute, Center for Electron Microscopy, School of Physics and Optoelectronics, State Key Laboratory of Luminescent Materials and Devices, Guangdong-Hong Kong-Macao Joint Laboratory of Optoelectronic and Magnetic Functional Materials, South China University of Technology, Guangzhou 511442, China}

\author{Nikolai~S.~Kiselev}
\email{n.kiselev@fz-juelich.de}
\affiliation{Peter Gr\"unberg Institute and Institute for Advanced Simulation, Forschungszentrum J\"ulich and JARA, 52425 J\"ulich, Germany}

\author{Filipp~N.~Rybakov}
\affiliation{Department of Physics and Astronomy, Uppsala University, Box-516, Uppsala SE-751 20, Sweden}

\author{Xiaodong Han}
\affiliation{Beijing Key Laboratory of Microstructure and Property of Advanced Materials, College of Materials Science and Engineering, Beijing University of Technology, Beijing 100124, China}
\affiliation{Department of Materials Science and Engineering, Southern University of Science and Technology, Shenzhen 518055, China}

\author{Stefan~Bl\"ugel}
\affiliation{Peter Gr\"unberg Institute and Institute for Advanced Simulation, Forschungszentrum J\"ulich and JARA, 52425 J\"ulich, Germany}

\author{Rafal~E.~Dunin-Borkowski}
\affiliation{Ernst Ruska-Centre for Microscopy and Spectroscopy with Electrons and Peter Gr\"unberg Institute, Forschungszentrum J\"ulich, 52425 J\"ulich, Germany}
\date{\today}
\maketitle

\textbf{
Magnetic skyrmions are topologically nontrivial spin configurations that possess particle-like properties. 
Earlier research was mainly focused on a specific type of skyrmion with topological charge $Q=-1$.
However, theoretical analyses of two-dimensional chiral magnets have predicted the existence of skyrmion bags -- solitons with arbitrary positive or negative topological charge.
Although such spin textures are metastable states, recent experimental observations have confirmed the stability of isolated skyrmion bags in a limited range of applied magnetic fields.
Here, by utilizing Lorentz transmission electron microscopy, we show the extraordinary stability of skyrmion bags in thin plates of B20-type FeGe. 
In particular, we show that skyrmion bags embedded within a skyrmion lattice remain stable even in zero or inverted external magnetic fields. 
A robust protocol for nucleating such embedded skyrmion bags is provided.
Our results agree perfectly with micromagnetic simulations and establish thin plates of cubic chiral magnets as a powerful platform for exploring a broad spectrum of topological magnetic solitons.
}

\vspace{0.5cm}

\begin{figure*}[ht] 
    \centering
    \includegraphics[width=18cm]{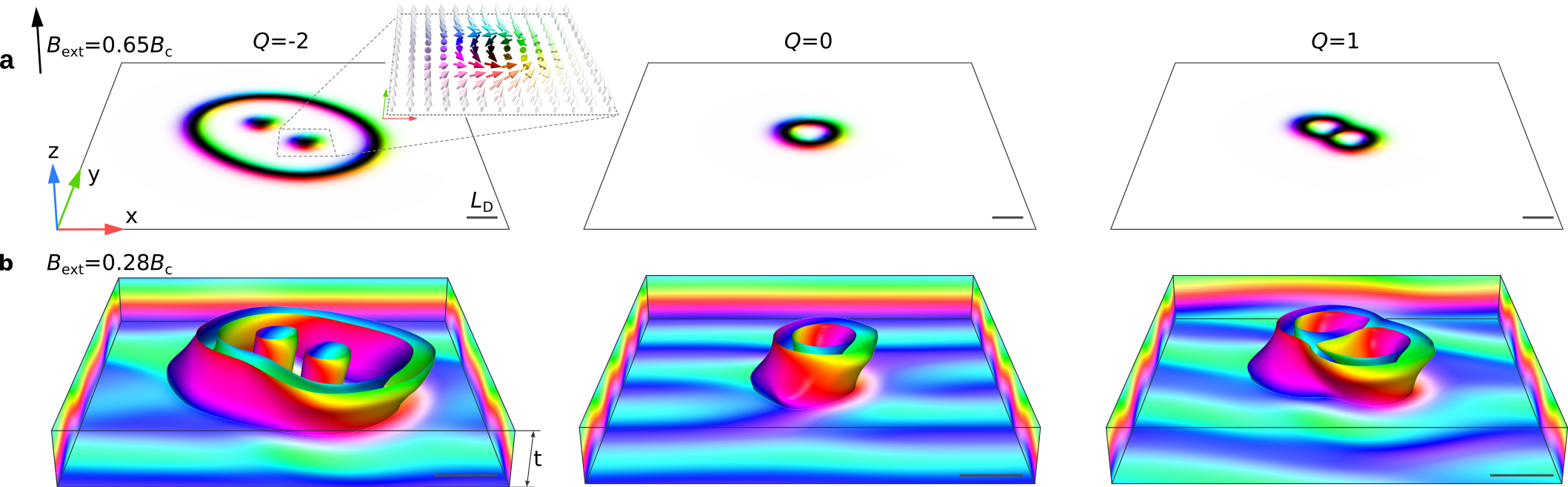}
    \caption{\small
    \textbf{Skyrmions of arbitrary topological charge in films of isotropic chiral magnets.}
    Horizontal panels \textbf{a} and \textbf{b} show representative skyrmion bags with topological charges ${Q=-2}$, $0$ and~$1$.
    \textbf{a} illustrates skyrmion bags in a 2D system.
    The magnetization field direction is depicted by means of a standard color code in all of the images, as explained in the inset.
    \textbf{b} illustrates skyrmion bags in a 3D system.
    The plate thickness $t$ is close to an equilibrium period of the chiral modulation $L_\mathrm{D}$ (${t = 0.85 L_\mathrm{D}}$).
    The magnetization field in \textbf{b} is visualized in the form of isosurfaces of ${m_z=0}$ and magnetization at the boundaries of the simulated domain.
    Demagnetizing fields are taken into account for a plate of finite thickness in \textbf{b} and are omitted for the 2D system in \textbf{a}.
    In both cases, the external magnetic field $B_\mathrm{ext}$ is perpendicular to the film. Its magnitude is given with respect to the critical field~$B_\mathrm{c}$. See Methods for details.
    The scale bar in all images corresponds to~${1\,L_\mathrm{D}}$.
    }
    \label{Fig_1}
\end{figure*}

Topological magnetic solitons~\cite{Kovalev_90} are stable configurations of magnetization fields that are characterized by nontrivial topology.
Similar to classical particles or particles in high energy physics, topological magnetic solitons can move and interact with each other in response to external stimuli, \emph{e.g.}, magnetic field~\cite{Tokura_20, Du_18}, temperature~\cite{Wang_20, Yu_21, Qin_22} and electric current~\cite{Jonietz_10, Yu_12, Wang_22}.
While experiments are typically carried out on three-dimensional (3D) samples of magnetic crystals, most of the magnetic solitons that have been observed belong to the class of two-dimensional (2D) topological solitons~\cite{Everschor_18, Gobel_21}.
The corresponding magnetic configurations are therefore only localized in two spatial directions, usually in the plane of the sample.
In the third direction, they extend through the sample to its physical edges.
In chiral magnets, an elementary 2D topological soliton represents a vortex, in which the magnetization twists by $\pi$ from its center to its periphery, as shown in the inset to Fig.~\ref{Fig_1}\textbf{a}.
In bulk samples, such $\pi$-vortices extend through the full sample thickness and are referred to as skyrmions.
It is common to refer to such filamentary magnetic textures as skyrmion strings~\cite{Birch_20, Seki_21, Zheng_21, Yu_22}.
The magnetic textures of such strings in any plane parallel to the sample surfaces are equivalent to each other up to trivial transformations.

For magnetic textures that are localized in 2D, the classifying group is the second homotopy group of the two-sphere.
This group is isomorphic to the group of integers with respect to addition: $\pi_2(\mathbb{S}^2, \mathbf{m}_0)=\mathbb{Z}$, where $\mathbf{m}_0$ is the base point~\cite{Hu_1959}.
A localized continuous magnetic texture can thereby be attributed to an integer number, which is usually termed the skyrmion topological charge and can be calculated using the following invariant:
\begin{equation}
    Q = \frac{1}{4\pi}\int_{\Omega}  \mathbf{m}\cdot \left[\partial_x\mathbf{m}\times\partial_y\mathbf{m}\right] \mathrm{d}x\mathrm{d}y, \label{Q}
\end{equation}
where $\mathbf{m}(x,y)=\mathbf{M}/|\mathbf{M}|$ is the unit vector field of magnetization and $\Omega$ is the skyrmion localization area, such that at its boundary $\partial\Omega$ the magnetization field ${\mathbf{m}(\partial\Omega)=\mathbf{m}_0}$.
Following the standard convention, the right-handed Cartesian coordinate system $xyz$ is chosen so that ${\mathbf{m}_0\cdot\mathbf{e}_z >0}$. For additional details, see Ref.~\cite{Zheng_23}.

Two magnetic textures with the same topological charge are topologically equivalent, meaning that they can be transformed continuously into each other.
By definition, magnetic configurations with $Q\neq0$ cannot be transformed continuously into a ferromagnetic state without the formation of singularities, which are typically associated with high energy states, suggesting that the decay of a topological soliton requires an energy barrier to be overcome.
The experimentally observed collapse and nucleation of skyrmions indicate that, at certain conditions, the system can overcome this energy barrier~\cite{Muckel_21}.
Thereby, the stability of magnetic solitons, besides the above topological arguments, is also determined by the interplay of competing magnetic interactions.

Here, we study magnetic skyrmions in thin plates of isotropic chiral magnets that are stabilized by a competition between Heisenberg ferromagnetic exchange and the chiral Dzyaloshinskii-Moriya interaction (DMI)~\cite{Dzyaloshinskii, Moriya}, in the presence of external magnetic fields applied perpendicular to the plates~\cite{Bogdanov_89}.

Axially symmetric solutions, or so-called $k\pi$-skyrmions~\cite{Bogdanov_99}, are known to have topological charges of $Q=0$ for even $k$ and $Q=-1$ for odd $k$.
Most earlier experimental studies of B20-type crystals of chiral magnets have been devoted to the static, dynamic and transport properties of $\pi$-skyrmions with $Q=-1$~\cite{Tokura_20, Nagaosa_13, Muhlbauer_09}.
However, recent theoretical studies of 2D models of chiral magnets have predicted that there is an infinite diversity of skyrmions with arbitrary positive and negative topological charges~\cite{Rybakov_19, Foster_19}.
We refer to these configurations as \textit{skyrmion bags}, following established terminology in the literature.

The ground state of an isotropic chiral magnet is a homochiral spin spiral.
In the absence of an external magnetic field, this spiral has a period $L_\mathrm{D}=4\pi\mathcal{A}/\mathcal{D}$, which is defined by the ratio of the Heisenberg exchange stiffness $\mathcal{A}$ and the DMI constant $\mathcal{D}$.
In this case, the stability of the skyrmions requires the presence of an external magnetic field.
Representative examples of skyrmion bags with $Q=-2$, $0$ and $1$ in a 2D system are shown in Fig.~\ref{Fig_1}\textbf{a}.
Figure~\ref{Fig_1}\textbf{b} shows skyrmion bags with the same topological charges obtained as stable solutions in a 3D system, in which the plate thickness $t$ is assumed to be close to the period of the helical modulations $L_\mathrm{D}$.
Details about the model and micromagnetic simulations are given in the Methods section.
Due to the isotropic properties of ferromagnetic exchange and DMI, these magnetic configurations show additional modulations along the thickness of the plate.
As a result, isosurfaces of the skyrmion bags show distortions and twists along the $z$-axis.
The magnetization field around a skyrmion bag is not a ferromagnetic state with $\mathbf{m} \parallel \mathbf{e}_z$, as in the 2D case. 
Instead, here the magnetization field takes the form of a conical spin spiral, whose $k$-vector is parallel to the external magnetic field $\mathbf{B}_\mathrm{ext}$.
As the topological charge remains fixed in any $xy$-plane, such configurations can be identified as quasi-2D skyrmion bags.

\begin{figure*}[ht] 
    \centering
    \includegraphics[width=17.5cm]{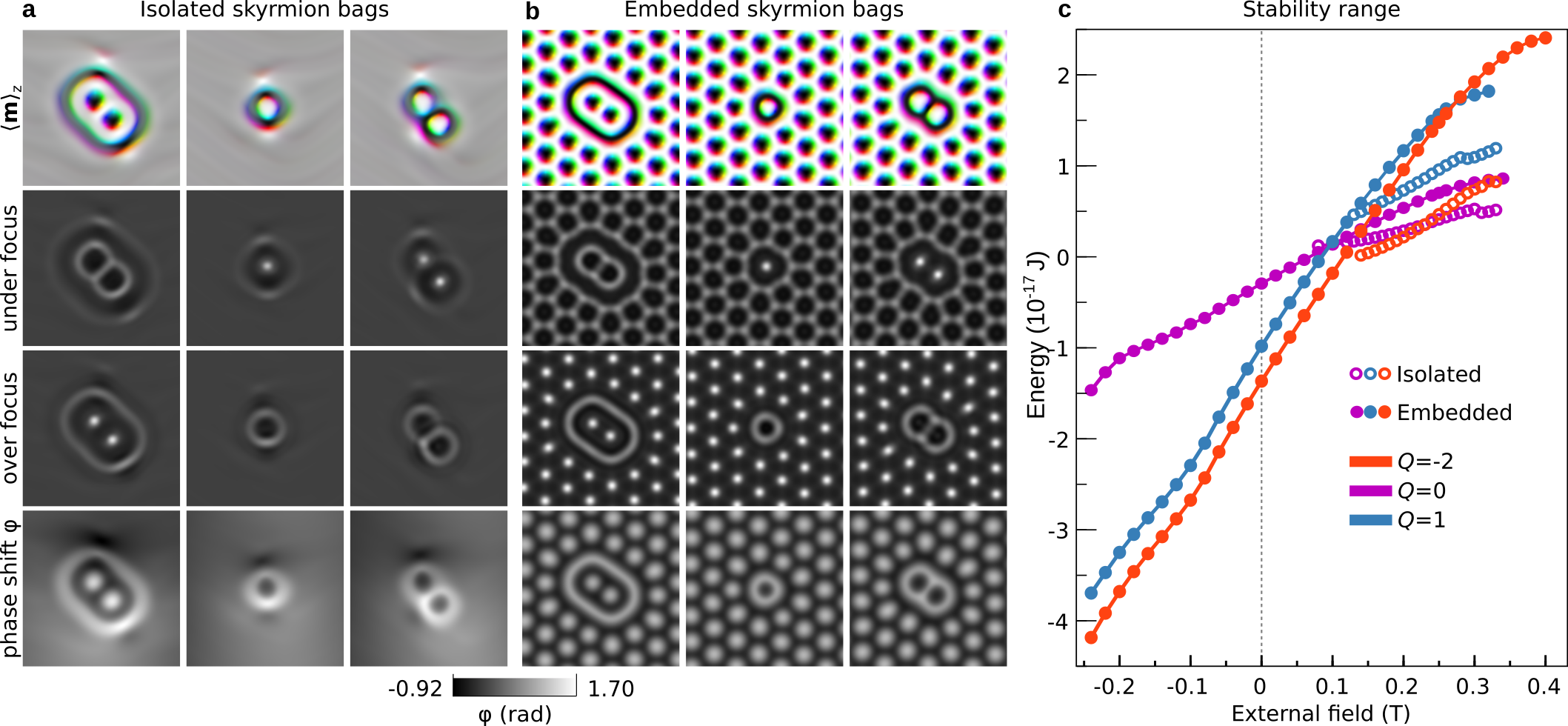}
    \caption{\small
    \textbf{Micromagnetic simulations of isolated and embedded skyrmion bags.}
    \textbf{a} corresponds to solutions for isolated skyrmion bags, while \textbf{b} shows skyrmion bags embedded in skyrmion lattices.
    The first row shows the magnetization field averaged over the thickness of the plate for skyrmion bags with $Q~=~-2$, $0$ and $1$, as shown in Fig.~\ref{Fig_1}b.
    Here, we use a specific color code~\cite{Savchenko_22}, with convergence to a gray color when $|\langle \mathbf{m} \rangle _z| \rightarrow 0$.
    The second and third rows show corresponding calculated over-focus and under-focus Lorentz TEM images.
    The fourth row shows the calculated phase shift, $\varphi$, which would be recorded using off-axis electron holography.
    All phase shift images are shown on the same scale.
    \textbf{c} plots the energies of isolated and embedded skyrmion bags as a function of the perpendicular applied magnetic field.
    The energies of the isolated solutions are given with respect to the energy of the conical phase, while the energies of the embedded solutions are given with respect to the energy of a regular skyrmion lattice.
    The isolated skyrmion bags only remain stable over a finite range of fields applied in the positive direction.
    The embedded skyrmion bags are stable over a wider range of applied fields, including inverted fields.
    }
    \label{Fig_2}
\end{figure*}

Experimental observations and current-driven motion of skyrmion bags with positive topological charge ${0\leq Q\leq 52}$ were recently reported in chiral magnets and referred to as \textit{skyrmion bundles}~\cite{Tang_21, Tang_23}.
The topologically trivial soliton with $Q=0$ depicted in the central column of Fig.~\ref{Fig_1} was also observed experimentally earlier~\cite{Tang_21, Zheng_17, Zheng_22}.  
%
Recently, skyrmion bags of diverse topological charges were also reported in the system without chiral interaction, e.g., in van der Waals ferromagnets~\cite{Powalla_23} where they were referred to as \textit{composite skyrmions} and in hexaferrite~\cite{Tang_23:2}.
However, the observation of such skyrmion bags in chiral magnets has proved to be more challenging.
It was recently shown that in thick films of an isotropic chiral magnet with $t \gtrsim 2L_\mathrm{D}$ the outer rings of skyrmion bags 
tend to transform into hopfion rings~\cite{Zheng_23}.
Moreover, in Ref.~\cite{Zheng_23} it was shown that 3D magnetic textures should be characterized by a more general homotopy group, in which the corresponding invariant is defined by the \textit{skyrmion-hopfion} topological charge represented by a pair of integers $(Q, H)$.
Formally, magnetic configurations with $(Q\neq0, H=0)$ should be attributed to 2D topological solitons, whereas configurations with $(Q, H\neq0)$ should be attributed to 3D topological solitons.
Representative examples of skyrmion bags shown in Fig.~\ref{Fig_1} belong to the class of 2D topological solitons.
The outer ring of a skyrmion bag with $Q=-2$ shown in Fig.~\ref{Fig_1}\textbf{b} can only transform into a hopfion ring in a relatively thick film.
Therefore, a plate with a thickness in the range $0< t\lesssim L_\mathrm{D}$ (for B20-type FeGe, $L_\mathrm{D}=70$~nm) is more suitable for the observation of skyrmion bags. 
The fabrication of such plates with a thickness below 100~nm using the standard method of focused ion beam milling is challenging.
In the present work, we describe the experimental observation of magnetic skyrmion bags with negative topological charge embedded in a skyrmion lattice in a 70-nm-thick FeGe plate.
We also provide a reliable approach for their nucleation, based on the experimental diagram of magnetic states in such a thin sample explored in our earlier study~\cite{Zheng_22}.

\noindent \textbf{Results} 

\noindent \textbf{Micromagnetic simulations}.
Figure~\ref{Fig_2} shows micromagnetic simulations, which demonstrate that the range of external magnetic fields in which isolated skyrmion bags remain stable is relatively narrow.
When $B_\mathrm{ext}$ is reduced below $\approx 120$~mT, the isolated skyrmion bags exhibit an instability with respect to elongation [Supplementary Fig.~\ref{FigS1}\textbf{a}-\textbf{c}].
When $B_\mathrm{ext}$ is increased above $340$~mT, the isolated skyrmion bags become unstable with respect to topological transitions.
In our simulations, configurations with $Q=0$ and $1$ transform into chiral bobbers [Supplementary Fig.~\ref{FigS1}\textbf{d}], whereas skyrmion bags with $Q=-2$ transform into pairs of skyrmions through collapse of their outer rings [Supplementary Fig.~\ref{FigS1}\textbf{e}].

In contrast, we find that skyrmion bags that are embedded in skyrmion lattices remain stable over very wide ranges of applied field, even when the directions of these fields are inverted.
In our calculations, we assume periodic boundary conditions in the plane of the plate and an extended system with a fixed skyrmion density $\sim 150$ $\mu$m$^{-2}$, in order to approximate the samples of confined geometry that are used in our experiments.
Our results show that a variation in skyrmion density does not change the skyrmion bag stability range significantly. 
Most importantly, the embedded skyrmion bags remain stable even in negative fields of up to $B_\mathrm{ext}=-240$~mT.
For $B_\mathrm{ext}<-240$~mT, even though the structure of the skyrmion lattice undergoes irreversible changes, the skyrmion bags can still be identified when the magnetic field is switched back to the positive direction. 
As a result, even $B_\mathrm{ext}=-240$~mT is not a strict critical field for embedded skyrmions, but can be regarded as an approximate lower bound.
Such extraordinary stability makes embedded skyrmion bags more accessible for experimental observations when compared to isolated skyrmion bags.

In positive fields, the embedded skyrmion bags exhibit an instability at nearly the same field of $\sim 350$~mT as isolated skyrmion bags.
In our micromagnetic simulations, at $B_\mathrm{ext}>320$~mT a skyrmion bag with $Q=1$ transitions into a skyrmionium with $Q=0$, which itself transforms into an ordinary skyrmion with $Q=-1$ at $B_\mathrm{ext}>350$~mT. 
The skyrmion bag with $Q=-2$ turns out to be the most stable solution, which then collapses into an ordinary skyrmion at $B_\mathrm{ext}>400$~mT.
Although the energies of the skyrmion bags are provided in Fig.~\ref{Fig_2}\textbf{c} to illustrate the difference in stability range between isolated and embedded skyrmion bags, it should be noted that a direct comparison of energies may be misleading since the backgrounds for isolated and embedded skyrmions are different.

\begin{figure}[h] 
    \centering
    \includegraphics[width=8.5cm]{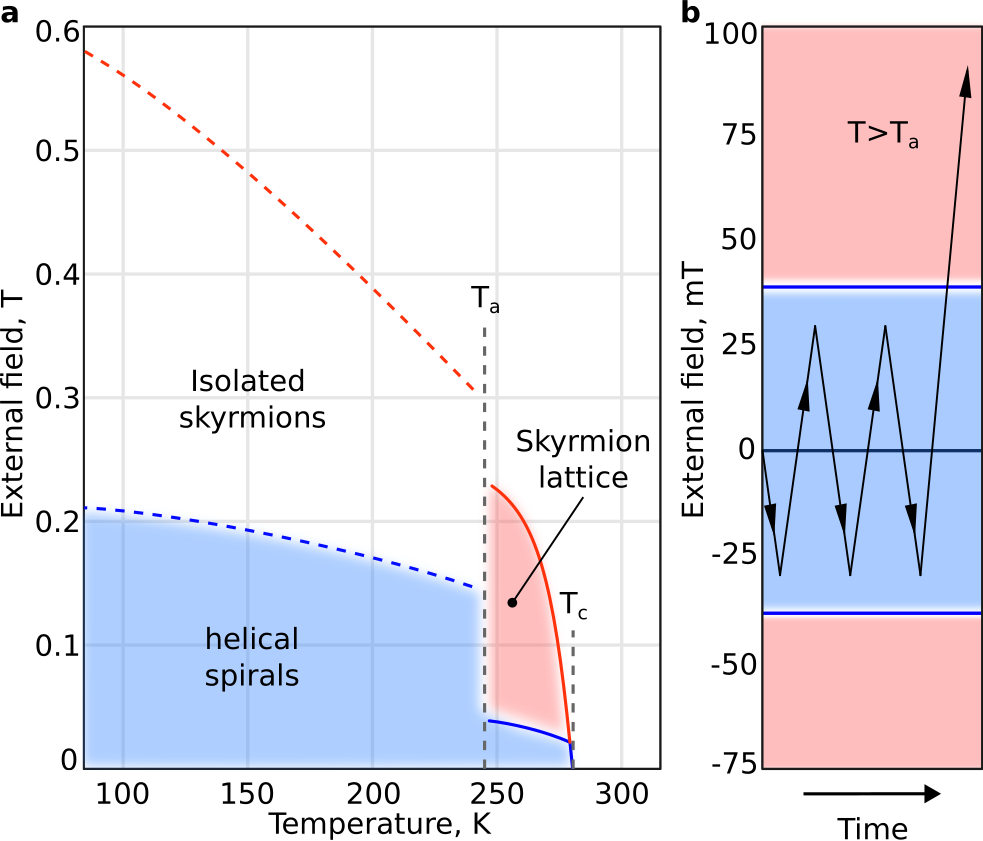}
    \caption{\small
    \textbf{Temperature - applied magnetic field diagram of magnetic states in a 70-nm-thick FeGe plate.}
    \textbf{a} shows a diagram of magnetic states reproduced using data from Ref.~\cite{Zheng_22}.  
    We estimate the activation temperature $T_\mathrm{a}=245$~K and the critical temperature $T_\mathrm{c}=278$~K.
    For $T~<~T_\mathrm{a}$, the dashed blue line marks the instability of the helical spin spiral state with respect to a transition to individual skyrmions. 
    The dashed red line corresponds to collapse of skyrmions.
    For $T_\mathrm{a}<T<T_\mathrm{c}$, the solid blue line marks an abrupt transition between a helical spin spiral state and a regular skyrmion lattice. The solid red line corresponds to the nearly simultaneous collapse of all skyrmions.
    \textbf{b} shows a schematic representation of our embedded skyrmion bag nucleation protocol at $T>T_\mathrm{a}$. 
    We first apply several cycles of a lower field between negative and positive directions.
    The applied field is then gradually increased above a value that induces a transition from helical spin spirals to a skyrmion lattice.
    }
    \label{Fig_PD}
\end{figure}

\noindent \textbf{Experimental stability diagram of magnetic states in a thin FeGe plate.}
We studied a thin FeGe plate with a lateral dimension of 1~$\mu$m~$\times$~1~$\mu$m and a thickness of $70$~nm, which corresponds to $1L_\mathrm{D}$. 
This sample was used in our previous work~\cite{Zheng_22}, where we estimated the stability diagram of magnetic states, which is reproduced in Fig.~\ref{Fig_PD}\textbf{a}.
The blue and red dashed lines were estimated in increasing magnetic fields.
Below the dashed lines, one can observe either isolated skyrmions or helical spirals.

The diagram shows that, when the temperature increases above the so-called activation temperature $T_\mathrm{a}$, the system undergoes an abrupt transition into a skyrmion lattice phase.
This transition can be induced by varying the applied magnetic field or temperature.
At a sample temperature in the range $T_\mathrm{a}\lesssim T\lesssim T_\mathrm{c}$ and an external field in the range $40$ mT $\lesssim B_\mathrm{ext}\lesssim 220$ mT, we always observe the appearance of a skyrmion lattice.
The protocol for the nucleation of embedded skyrmion bags that we introduce below relies on the abrupt transition between helical spirals and a skyrmion lattice induced by an external magnetic field.

\begin{figure*}[ht] 
    \centering
    \includegraphics[width=17.5cm]{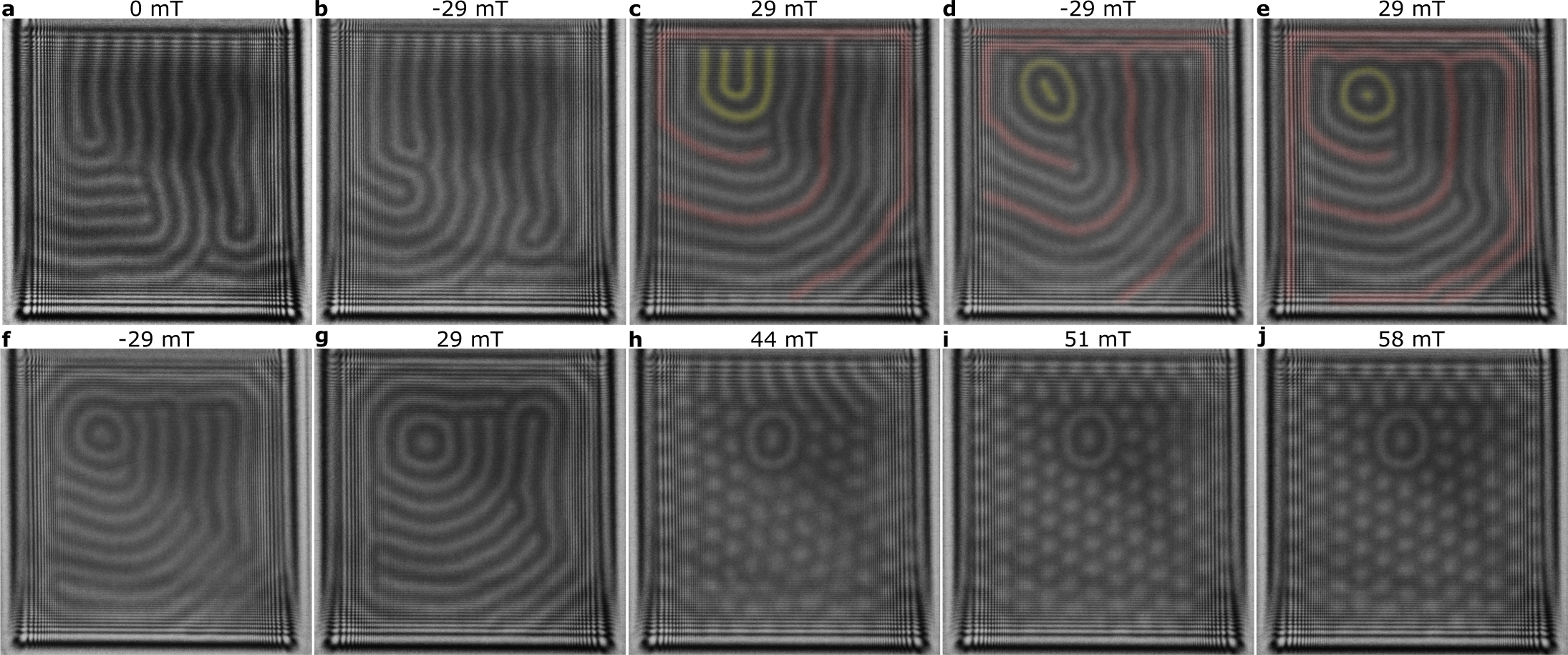}
    \caption{\small
    \textbf{Nucleation protocol for skyrmion bags in a 70-nm-thick FeGe plate.}
    \textbf{a} shows a random initial magnetic state in zero magnetic field.
    We chose a magnetic field of $\sim$~30~mT for the field swapping process.
    \textbf{b}-\textbf{c} show states in the first swapping cycle between $-29$ and $+29$~mT. 
    \textbf{c} shows the formation of a helical spiral near the edge (red line). 
    Two short spirals (yellow lines) are pushed and turned into a skyrmion bag, with a skyrmion inside them in the second cycle, as shown in \textbf{d} and \textbf{e}. The skyrmion bag is located in the background of the helical spirals, with a topological charge of $-1$.
    Two further helical modulations are created continuously from the edge of the sample with more cycles of field swapping, as shown in \textbf{d}-\textbf{e} and \textbf{f}-\textbf{g}.
    By following the red line, these newly-formed helical spirals are pushed gradually towards the inner area of the sample. 
    On increasing the field further, such helical spirals transition into skyrmions, resulting in skyrmion bags embedded in the skyrmion lattice, as shown in \textbf{h}, \textbf{i} and \textbf{j}.
    The frames show over-focus Lorentz TEM images recorded at $T$~=~250~K. The defocus distance is 800~$\mu$m. The value of the applied magnetic field is indicated above each frame.
    Supplementary Fig.~\ref{FigS-nucleation} shows another example of skyrmion bag nucleation in the sample.
    Supplementary Fig.~\ref{FigS-t180-nucleation} shows skyrmion bag nucleation in a 180-nm-thick FeGe plate. 
    }
    \label{Fig_nucleation}
\end{figure*}

\begin{figure*}[ht] 
    \centering
    \includegraphics[width=17.5cm]{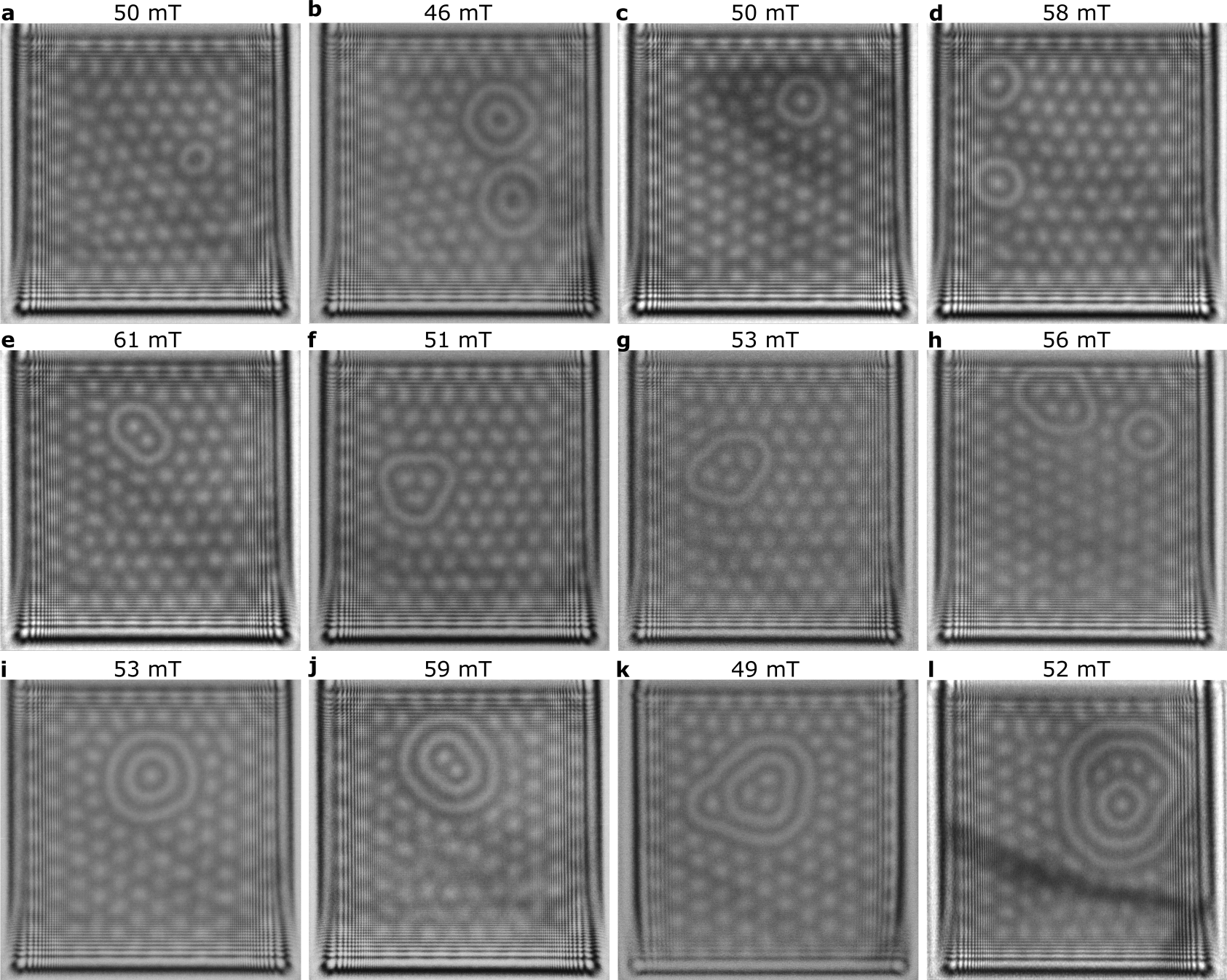}
    \caption{\small
    \textbf{Representative examples of skyrmion bags with negative topological charge in a 70-nm-thick FeGe plate.}
    \textbf{a} shows a single embedded skyrmionium.
    \textbf{b} shows  two so-called $4\pi$-skyrmions with $Q=0$.
    \textbf{c} shows a single $3\pi$-skyrmion, \emph{i.e.}, a skyrmion bag with $Q=-1$.
    \textbf{d} shows a pair of $3\pi$-skyrmions.
    \textbf{e}, \textbf{f} and \textbf{g} show embedded skyrmions with two, three and four skyrmions inside outer rings and topological charge $Q=-2$, $-3$ and $-4$, respectively.
    \textbf{h} shows the coexistence of two skyrmion bags with $Q=-4$ and $-1$.
    \textbf{i} shows an embedded $5\pi$-skyrmion.
    \textbf{j}, \textbf{k} and \textbf{l} show non-axially-symmetric skyrmion bags with multiple rings and topological charge $Q=-2$, $-4$ and $-5$, respectively.
    }
    \label{Fig_collection}
\end{figure*}

\noindent\textbf{Protocol for skyrmion bag nucleation.} 
Figure~\ref{Fig_nucleation} shows a sequence of over-focused Lorentz TEM images, which illustrate a protocol for the reliable nucleation of skyrmion bags.
The images were recorded at sample temperature of $T=250$~K, which is above the activation temperature.
At this temperature, the transition into a skyrmion lattice occurs at $\sim 39$~mT. 
The magnitude of the magnetic field applied in the positive and negative directions should therefore be lower than this value.
We estimated the optimal magnitude of the magnetic field to be $\lesssim~30$~mT.

We start from a helical spiral state in zero magnetic field at a temperature of $T>T_\mathrm{a}$, as shown in Fig.~\ref{Fig_nucleation}\textbf{a}.
We first apply a few cycles of varying magnetic field, which is always oriented perpendicular to the plate but changes in direction from positive to negative, as shown schematically in Fig.~\ref{Fig_PD}\textbf{b}.
The aim of this step is to create closed loops of helical spin spirals.
Figure~\ref{Fig_nucleation}\textbf{c}, \textbf{d} and \textbf{e} shows the formation of helical spirals (marked by red lines) near the edges, while two helical spirals (marked by yellow lines) in the interior of the sample form a closed loop after the first and second field swapping cycles. A skyrmion bag with $Q=-1$ is embedded in the helical spirals.
We then increase the field gradually to induce a transition from the helical spirals to a skyrmion lattice state, as shown in Fig.~\ref{Fig_nucleation}\textbf{h}, \textbf{i} and \textbf{j}.

With more field-swapping cycles, we observe that closed loops can also be formed from the edges, leading to a diversity of skyrmion bags embedded within the helical spirals and skyrmion lattice.
In this way, the number of skyrmions inside the bag and the number of closed loops can be controlled.
Furthermore, the initial configuration of the system influences the diversity of the skyrmion bags.
For instance, Supplementary Fig.~\ref{FigS-nucleation} illustrates the above protocol for an initial configuration of an ideal helical spin spiral. 
This protocol is conceptually identical to that used for the nucleation of skyrmion-antiskyrmion pairs~\cite{Zheng_22} and hopfion rings~\cite{Zheng_23}.
However, here we apply it at elevated temperature.

\noindent\textbf{Diversity of embedded skyrmion bags}.
By following the above protocol, we obtained a wide diversity of skyrmion bags embedded in skyrmion lattices. 
Representative examples of skyrmion bags with different topological charge are shown in Fig.~\ref{Fig_collection}.
Further examples of negative skyrmion bags are provided in Supplementary Fig.~\ref{FigS-collections}.

Positive skyrmion bags with $Q>0$ were not observed in our experiments.
Merging of skyrmion bags was also not observed.
Even when there were several skyrmion bags in the system, as shown in Fig.~\ref{Fig_collection}\textbf{b}, \textbf{d} and \textbf{h}, we never observed merging of their outer rings.

The shape and position of a skyrmion bag were found to depend not only on the number of field-swapping cycles, but also on the angle of the applied magnetic field.
In general, the use of a tilted magnetic field breaks the symmetry of the system by making specific directions of magnetization in the plane more favorable than others, resulting in an obstacle to the formation of closed loops of helical spirals from the sample edges.
The tilt angle of the magnetic field was therefore always kept below 5$^\circ$ in our experiments.

\noindent\textbf{Skyrmion bags in inverted magnetic fields.}
In order to verify the prediction of the micromagnetic model about the stability of skyrmion bags in inverted magnetic fields, we performed dedicated experiments at reduced temperature, since the micromagnetic constants used in the simulations were adapted for the low temperature regime.
We first nucleated skyrmion bags by following the above protocol at a higher temperature of 250~K and then cooled the sample to 95~K.
Representative examples of the field evolution of skyrmion bags with topological charges of $Q=0$, $-1$ and $-2$ in negative fields are shown in Supplementary Figs~\ref{FigS-flip1}, \ref{FigS-flip2} and \ref{FigS-flip3}, respectively.
All of the skyrmion bags exhibit topological instability in applied magnetic fields of between $-150$ and $-200$~mT, but are otherwise in good agreement with the micromagnetic simulations shown in Fig.~\ref{Fig_2}\textbf{c}, in which skyrmion bags embedded in a skyrmion lattice are shown to be stable over a wide range of external magnetic fields in both directions.
This slight discrepancy between theory and experiment can be attributed to the confined sample geometry and to the fact that thermal fluctuations are ignored in the micromagnetic simulations.

An intriguing asymmetry is observed in the contrast of the skyrmions, especially in strong negative fields in Supplementary Figs~\ref{FigS-flip1}, \ref{FigS-flip2} and \ref{FigS-flip3}.
This behavior resembles an initial stage of a process that is referred to as \textit{turning skyrmion inside out}~\cite{Kuchkin_20}, in which a skyrmion gradually transforms into an antiskyrmion~\cite{Zheng_22}.
With increasing negative field, the number of spins pointing along the field increases gradually.
The negative direction then takes over the role of the base point $\mathbf{m}_0$, which, in turn, leads to a flip of the coordinate system in Eq.~\ref{Q}. 
A dedicated discussion of this intriguing transformation will be presented elsewhere.

\noindent \textbf{Discussion}

\noindent 
In order to distinguish the observed skyrmion bags from hopfion rings, we performed quantitative measurements using off-axis electron holography [Supplementary Fig.~\ref{Fig_comparison}]. 
The skyrmions and the outer rings of the skyrmion bags produced nearly identical recorded phase shifts, confirming that the rings penetrate through the thickness of the plate.
In contrast, the signal from a hopfion ring would be significantly weaker~\cite{Zheng_23}.

The protocol presented here only allows skyrmion bags embedded in a skyrmion lattice to be nucleated in very thin films of FeGe.
We attempted to reproduce this protocol in a 180-nm-thick FeGe plate of identical lateral size (1~$\mu$m$\times 1~\mu$m).
Snapshots of the system during field swapping cycles are provided in Supplementary Fig.~\ref{FigS-t180-nucleation}. 
Over seven cycles, helical spirals from the edges (marked in red) propagated gradually to the center of the sample and finally formed a single skyrmion.
This final configuration represents a nearly ideal set of nested closed loops of helical spirals.
The behaviour of the helical spirals nucleating at the edges of thicker and thinner films is, therefore, nearly identical.
By following this approach, one can obtain skyrmion bags embedded in the helical spirals, as shown in Supplementary Fig.~\ref{FigS-t180-bag}.
However, on further increasing the applied magnetic field, the texture undergoes a transition to a skyrmion lattice state, including the helical spirals and skyrmion bags.
A representative example of such a state, after increasing the field above the transition to a skyrmion lattice, is shown in Supplementary Fig.~\ref{FigS-t180-bag}\textbf{h}.
This behavior can be explained by the fact that, in a thick film, a skyrmion bag state is energetically higher than that of a hopfion ring.
In order to reduce its energy, the system therefore shrinks any closed ring into a hopfion ring.
However, at higher temperatures the hopfion ring collapses, leading to a state in which only skyrmions are present.

In conclusion, our results provide direct experimental evidence for the formation of magnetic skyrmion bags embedded in skyrmion lattices or helical spirals in thin plates of B20-type FeGe.
Micromagnetic simulations support our observations and show excellent agreement with the experimental results.
Magnetic skyrmion bags with tunable topological charge offer a powerful platform to study both the fundamental physics and the dynamical and topological properties of magnetic solitons.

\renewcommand{\bibname}{Literary works}
%



\vspace{5mm}

\textbf{Acknowledgements.}
The authors are grateful for funding from the European Research Council under the European Union's Horizon 2020 Research and Innovation Programme (Grant No.~856538 - project ``3D MAGiC'') and to the Deutsche Forschungsgemeinschaft (Project-ID 405553726 – TRR~270; Priority Programme SPP 2137; Project No.~403502830).
L.Y. acknowledges financial support from the Helmholtz-OCPC Postdoc-Program from the Helmholtz Association and the Office of China Postdoc Council (Grant No. ZD2019018) and the National Natural Science Foundation of China (Grant No. 52204379).
F.N.R. acknowledges support from the Swedish Research Council (2023-04899).
N.S.K. acknowledges financial support from the Deutsche Forschungsgemeinschaft through SPP 2137 ``Skyrmionics" Grant No. KI 2078/1-1.
F.Z. acknowledges financial support from the Fundamental Research Funds for the Central Universities, the National Natural Science Fund for Excellent Young Scientists Fund Program (Overseas) and for the General Program (Grant No. 52373226).

\textbf{Author contributions.}
F.Z. and N.S.K. conceived the project and designed the experiments.
L.Y. performed the TEM experiments and data analysis.
A. S. and N.S.K. performed micromagnetic simulations, with assistance from F.N.R.
L.Y., A.S., N.S.K. and F.Z. prepared the manuscript.
All of the authors discussed the results and contributed to the final manuscript.

\textbf{Data availability.}
All data are available from the corresponding authors upon reasonable request.


\textbf{Competing interests.}
The authors declare no competing interests.

\newpage

\vspace{5mm}

\section*{Methods}

\textbf{Magnetic imaging in the TEM.}
TEM samples were prepared from a single crystal of B20-type FeGe using a focused ion beam workstation and a lift-out method~\cite{Du_15, Zheng_18}.
Fresnel defocus imaging and off-axis electron holography were performed at 300~kV in an FEI Titan 60-300 TEM equipped with an electrostatic biprism. 
The microscope was operated in the Lorentz mode with the sample either in magnetic-field-free conditions or in a pre-calibrated out-of-plane magnetic field applied using the conventional microscope objective lens.
A liquid-nitrogen-cooled specimen holder (Gatan model 636) was used to vary the sample temperature between 95 and 380~K. Fresnel defocus images and off-axis electron holograms were recorded using a 4k~$\times$~4k Gatan K2 IS direct electron counting detector. 
Multiple electron holograms, with a 4~s exposure time for each hologram, were recorded to improve the signal-to-noise ratio. Holograms were analyzed using a standard fast Fourier transform algorithm in Holoworks software (Gatan).
The magnetic induction maps that are shown in Supplementary Fig.~\ref{Fig_comparison} were obtained from the gradients of recorded magnetic phase images.

\textbf{Micromagnetic calculations.}
We used a micromagnetic model described by following energy density functional~\cite{Zheng_21, Fratta} :
\begin{align} 
\mathcal{E}\!=\!\int\limits_{V_\mathrm{m}}\!d\mathbf{r}\ 
&\mathcal{A}\sum\limits_{i=x,y,z} |\nabla m_i|^2 
+\mathcal{D}\,\mathbf{m}\!\cdot(\nabla\!\times\!\mathbf{m})
- M_\mathrm{s}\,\mathbf{m}\!\cdot\!\mathbf{B}_\mathrm{tot} + \nonumber\\
&+ \frac{1}{2\mu_0}\int\limits_{\mathbb{R}^3}\!d\mathbf{r}\ 
\sum\limits_{i=x,y,z}|\nabla A_{\mathrm{d}, i}|^2~,
\label{Ham_m}
\end{align}
where $\mathcal{A}$ and $\mathcal{D}$ are the Heisenberg exchange and DMI constants; $\mu_0$ is the vacuum permeability; ${\mathbf{m}(\mathbf{r})}=\mathbf{M}(\mathbf{r})/M_\text{s}$ is the unit vector field of the normalized magnetization, $M_\text{s}$ is the saturation magnetization, ${\mathbf{A}_\text{d}(\mathbf{r})}$ is the component of the magnetic vector potential induced by the magnetization~\cite{Fratta} and $V_\mathrm{m}$ is the volume of the calculation domain.
The total magnetic field $\mathbf{B}_\mathrm{tot}$ includes the external magnetic field $\mathbf{B}_\mathrm{ext}$
and the demagnetizing field produced by the sample~\cite{Zheng_21, Fratta}, according to the expression  $\mathbf{B}_\mathrm{tot}=\mathbf{B}_\mathrm{ext}+\nabla\!\times\!\mathbf{A}_\text{d}$.
We used the following parameters for FeGe~\cite{Zheng_18, Zheng_21, Zheng_22}:
$\mathcal{A}~=~4.75$~pJm$^{-1}$, $\mathcal{D}~=~0.853$~mJm$^{-2}$ and $M_\text{s}~=~384$~kAm$^{-1}$. 
Micromagnetic calculations based on energy minimization to find solutions of Eq.~\eqref{Ham_m}, as well as simulations of Lorentz TEM images and electron phase shift images, were performed using Excalibur software~\cite{Excalibur}.
Detailed descriptions of the methods are provided in Ref.~\cite{Zheng_21}. 

In order to estimate the stabilities of the skyrmion bags in Fig.~\ref{Fig_2}, we used calculation domains with sizes of 1.05~$\mu$m~$\times$~1.04~$\mu$m~$\times$~$70$~nm, on the assumption of periodic boundary conditions in the $xy$ plane. For the simulations of electron phase shift images shown in Supplementary Fig.~\ref{Fig_comparison}, the simulated domain had open boundary conditions in all directions.

\renewcommand{\bibname}{suppl references}


\setcounter{figure}{0}
\captionsetup[figure]{labelfont={bf},name={Supplementary Fig.},labelsep=period}

\begin{figure*}[ht] 
    \centering
    \includegraphics[width=15cm]{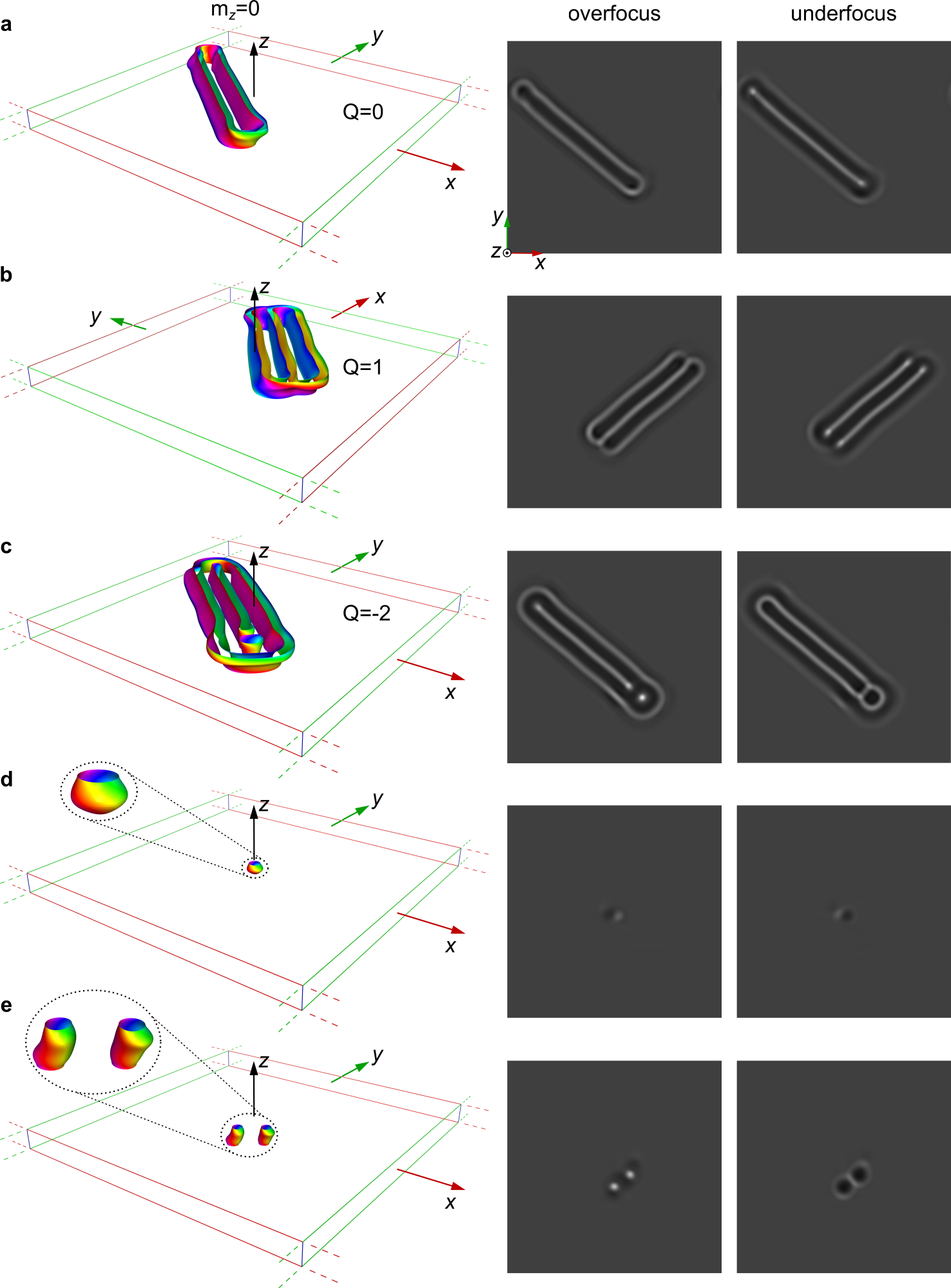}
    \caption{\small
    \textbf{Field evolution of skyrmion bags embedded in the conical phase.}
    \textbf{a}-\textbf{c} show skyrmion bags with $Q~=~0$, $+1$ and $-2$ embedded in the conical phase in a relatively lower magnetic field than the minimum of the stability range shown in Fig.~\ref{Fig_2}\textbf{c}.
    Elongation of the skyrmions implies an elliptical instability of the state.
    In a magnetic field higher than the stability range, skyrmion bags with $Q~=~0$ and $1$ transition to a chiral bobber (see \textbf{d}), whereas a skyrmion bag with $Q~=~-2$  transitions to two skyrmions (see \textbf{e}).
    The left column shows isosurfaces ($m_z~=~0$), while the middle and right columns show corresponding calculated over-focus and under-focus Lorentz TEM images for a defocus distance of 800~$\mu$m.
    }
    \label{FigS1}
\end{figure*}

\begin{figure*}[ht] 
    \centering
    \includegraphics[width=17.5cm]{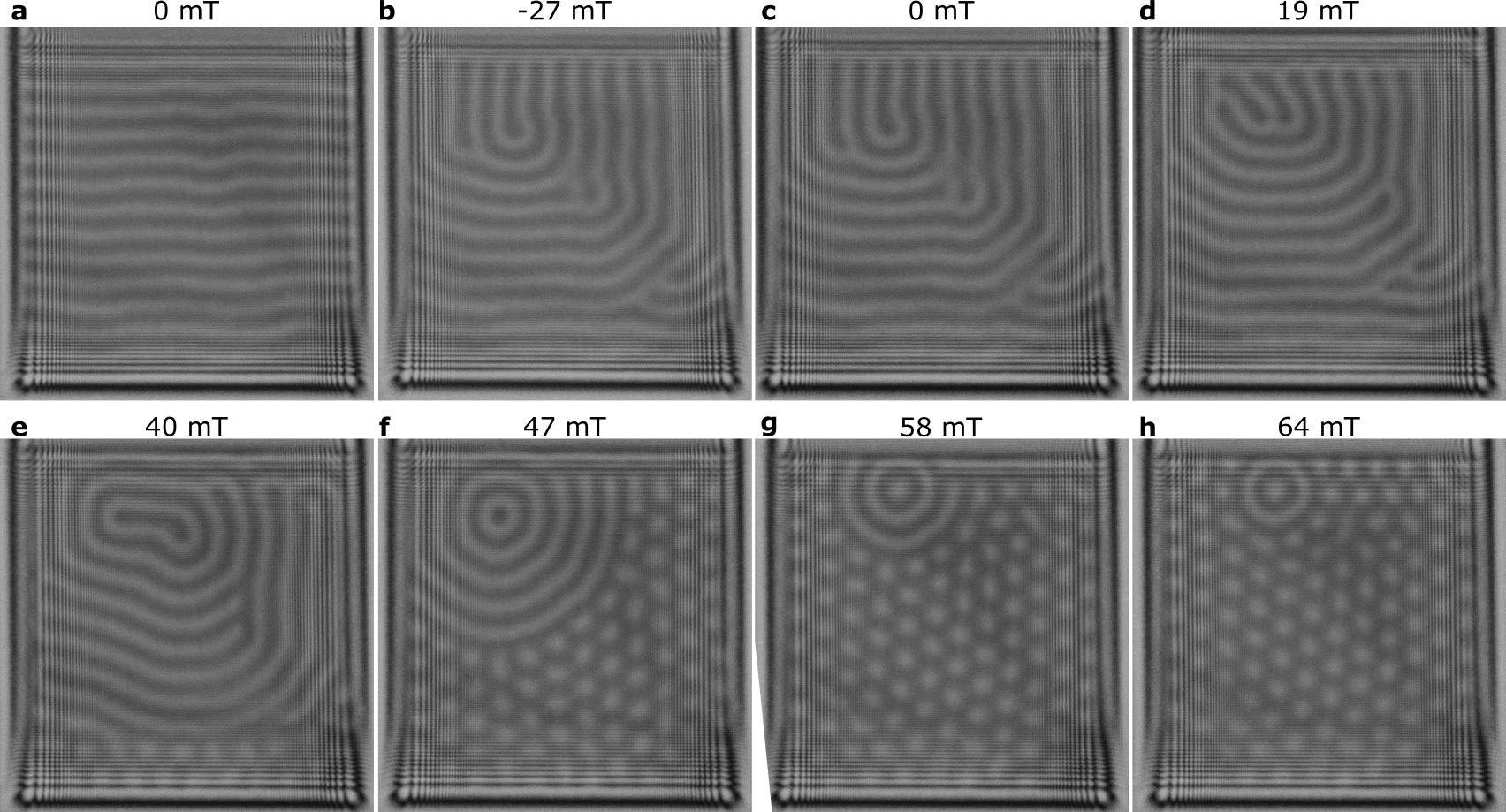}
    \caption{\small
    \textbf{Further example of the nucleation process of a skyrmion bag in a 70-nm-thick FeGe plate.}
    One cycle of magnetic field swapping results in the generation of helical spirals from the edge (see the upper edge in images \textbf{b}-\textbf{d}).
    On increasing the field, several closed loops of helical spirals are formed (see \textbf{e} and \textbf{f}).
    By increasing to a higher field, the inner closed loop of the helical spiral transitions into a skyrmion, leading to the formation of a skyrmion bag with a topological charge of -1 (see \textbf{g} and \textbf{h}).
    At 64~mT, a perfect skyrmion lattice is formed.
    The frames show over-focus Lorentz TEM images recorded at $T$~=~250~K. The defocus distance is 800~$\mu$m. The value of the applied magnetic field is indicated above each frame.
    }
    \label{FigS-nucleation}
\end{figure*}

\begin{figure*}[ht] 
    \centering
    \includegraphics[width=17.5cm]{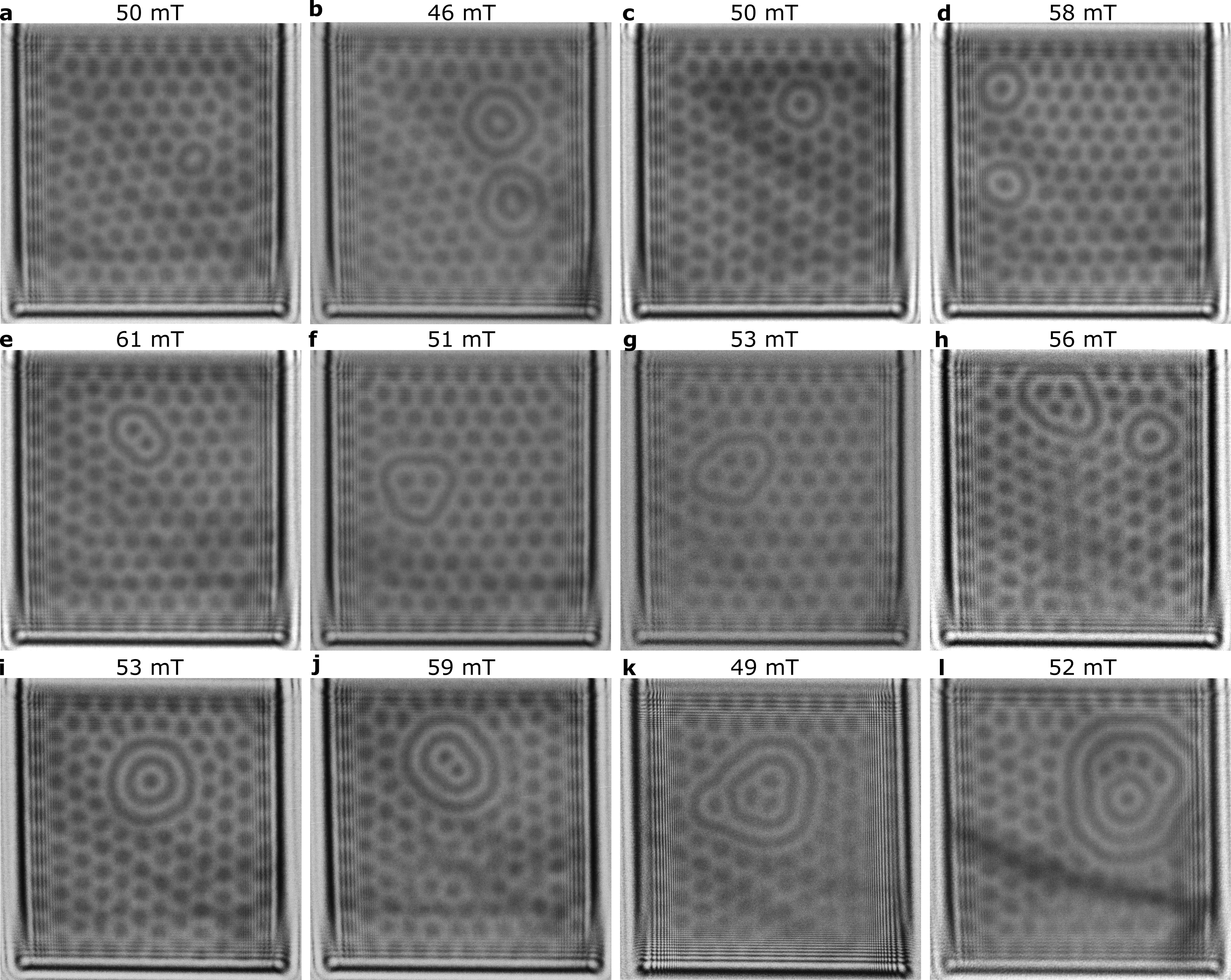}
    \caption{\small
    \textbf{Under-focus Lorentz TEM images of skyrmion bags corresponding to those shown in Fig.~\ref{Fig_collection}.}
    }
    \label{FigS-underfocus}
\end{figure*}

\begin{figure*}[ht] 
    \centering
    \includegraphics[width=17.5cm]{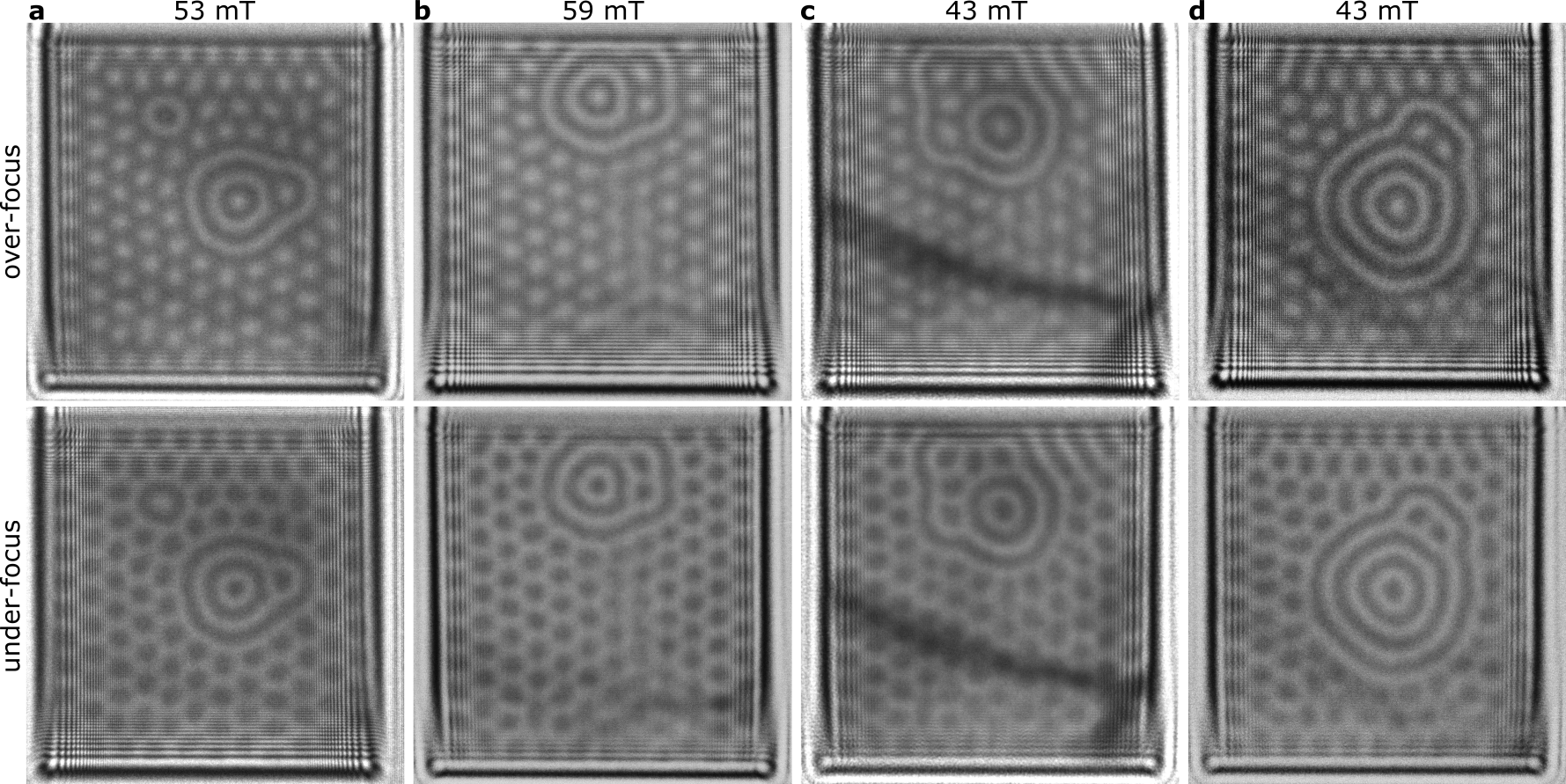}
    \caption{\small
    \textbf{Further variants of skyrmion bags with negative topological charge in a 70-nm-thick FeGe plate.}
    The upper and lower panels show over-focus and under-focus Lorentz TEM images of skyrmion bags, respectively. 
    \textbf{a} shows a configuration containing two separate skyrmion bags. The upper left one is a skyrmionium (also seen in Fig. \ref{Fig_collection}\textbf{a}-\textbf{b}). 
    The lower right one comprises two rings and two skyrmions, with one skyrmion located between the outer and inner rings and the other one in the inner ring. The total topological charge is $-2$.
    \textbf{b} shows a skyrmion bag with a total topological charge of $-3$. Two skyrmions are located between the outer and inner rings, while one skyrmion is located in the inner ring.
    \textbf{c} shows a skyrmion bag with a total topological charge of $-6$. Five skyrmions are located between the outer and inner rings, while one skyrmion is located in the inner ring.
    \textbf{d} shows a skyrmion bag containing three skyrmions and three rings, with a topological charge of $-3$. The outermost ring encloses two skyrmions, while the two inner rings surround one skyrmion.
    The defocus distance is 800 $\mu$m. The sample temperature is $T$~=~250~K.
    The value of the applied magnetic field is indicated above each frame.
    }
    \label{FigS-collections}
\end{figure*}

\begin{figure*}[ht] 
    \centering
    \includegraphics[width=17.5cm]{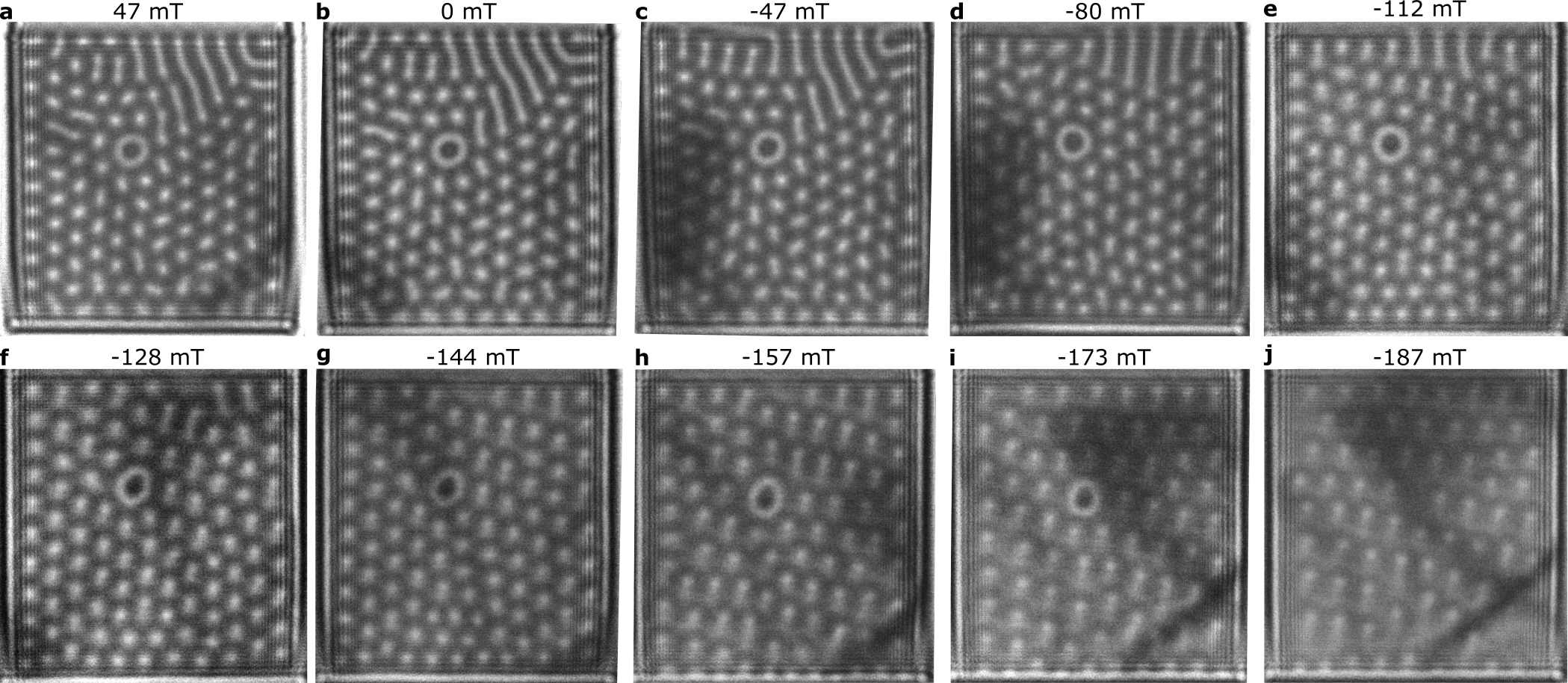}
    \caption{\small
    \textbf{Field evolution of a magnetic skyrmionium.}
    \textbf{a} shows a skyrmionium in a skyrmion lattice in a positive field of 47~mT.
    The magnetic field was then changed gradually towards the negative direction.
    The skyrmionium remained stable until a negative field of 187~mT (see \textbf{j}).
    Elongation of the skyrmionium and skyrmions was observed, in particular at higher negative fields (see \textbf{h}, \textbf{i} and \textbf{j}).
    The direction of elongation corresponds to the direction of the in-plane component of the applied magnetic field.
    The frames show over-focus Lorentz TEM images recorded at $T$~=~95~K. The defocus distance is 800~$\mu$m.
    The value of the applied magnetic field is indicated above each frame.
    }
    \label{FigS-flip1}
\end{figure*}

\begin{figure*}[ht] 
    \centering
    \includegraphics[width=17.5cm]{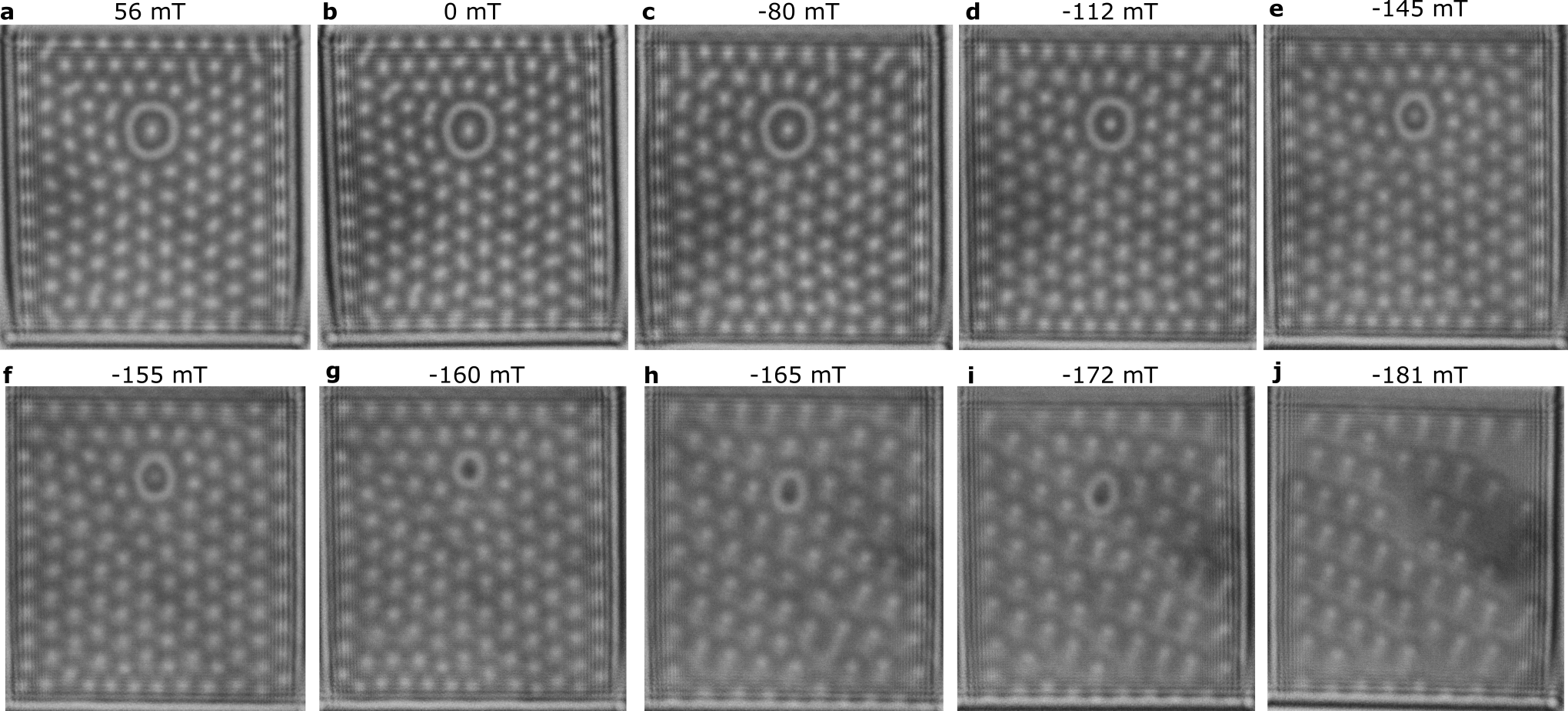}
    \caption{\small
    \textbf{Field evolution of a magnetic skyrmion bag with a topological charge of $-1$.}
    \textbf{a} shows a skyrmion bag with one skyrmion in a skyrmion lattice in a positive field of 56~mT.
    The magnetic field was then changed gradually towards the negative direction.
    The skyrmion bag remained stable until a negative field of 160~mT, before collapsing into a skyrmionium (see \textbf{g}).
    The size of the skyrmion bag decreased with increasing applied magnetic field (compare \textbf{a}, \textbf{e} and \textbf{f}).
    The skyrmionium collapsed in a negative field of 181~mT (see \textbf{j}).
    Elongation of the bag and the skyrmions was observed, in particular at higher negative fields (see \textbf{h}, \textbf{i} and \textbf{j}).
    The direction of elongation corresponds to the direction of the in-plane component of the applied magnetic field.
    The frames show over-focus Lorentz TEM images recorded at $T$~=~95~K. The defocus distance is 800~$\mu$m.
    The value of the applied magnetic field is indicated above each frame.
    }
    \label{FigS-flip2}
\end{figure*}

\begin{figure*}[ht] 
    \centering
    \includegraphics[width=17.5cm]{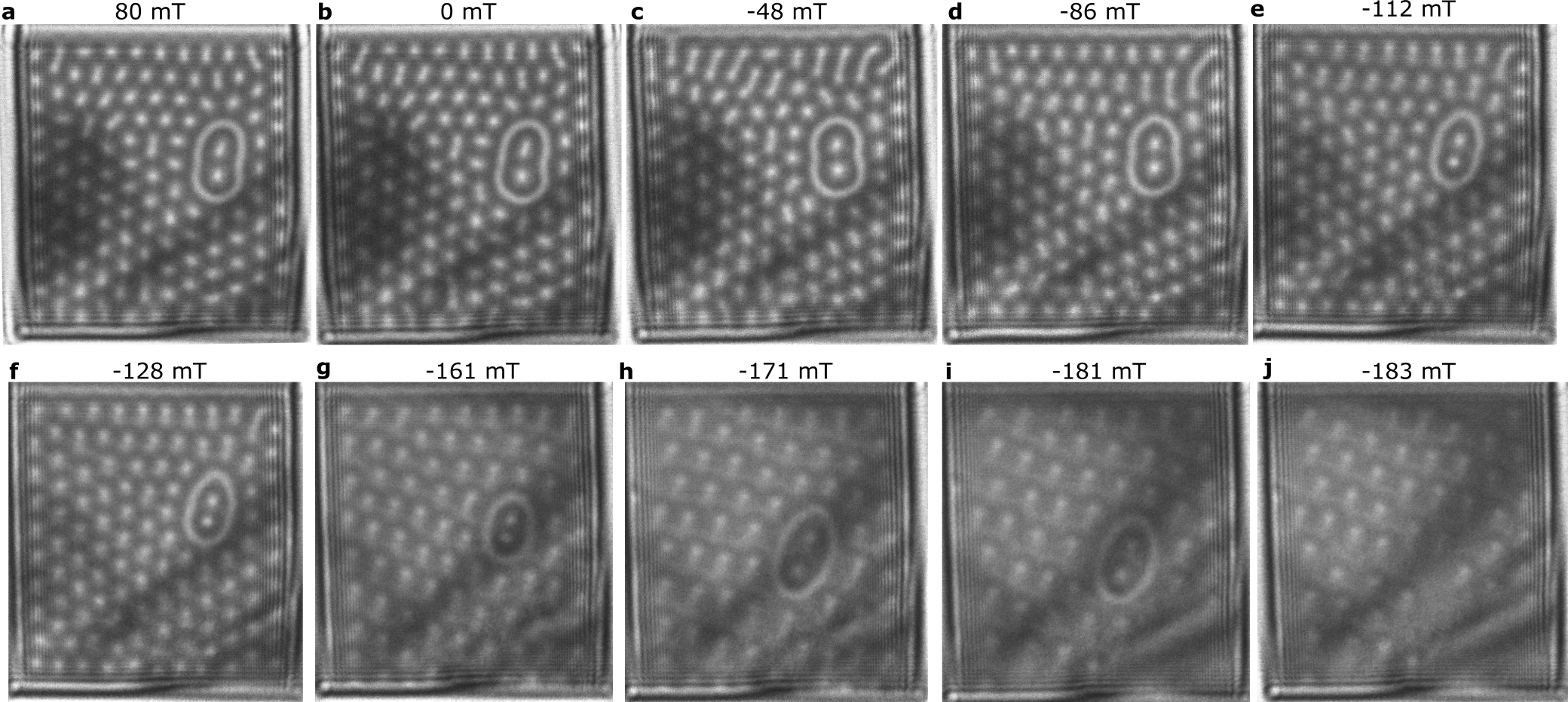}
    \caption{\small
    \textbf{Field evolution of a magnetic skyrmion bag with a topological charge of $-2$.}
    \textbf{a} shows a skyrmion bag with two skyrmions in a skyrmion lattice in a positive field of 80~mT.
    The magnetic field was then changed gradually towards the negative direction.
    The skyrmion bag remained stable until a negative field of 183~mT, before collapsing (see \textbf{j}).
    Elongation of the bag and the skyrmions was observed, in particular at higher negative fields (see \textbf{g}, \textbf{h}, \textbf{i} and \textbf{j}). 
   The direction of elongation corresponds to the direction of the in-plane component of the applied magnetic field.
The bag is smaller in \textbf{g} than in \textbf{h} and \textbf{i}.
    The frames show over-focus Lorentz TEM images recorded at $T$~=~95~K. The defocus distance is 800~$\mu$m.
    The value of the applied magnetic field is indicated above each frame.
    }
    \label{FigS-flip3}
\end{figure*}



\begin{figure*}[ht] 
    \centering
    \includegraphics[width=17.5cm]{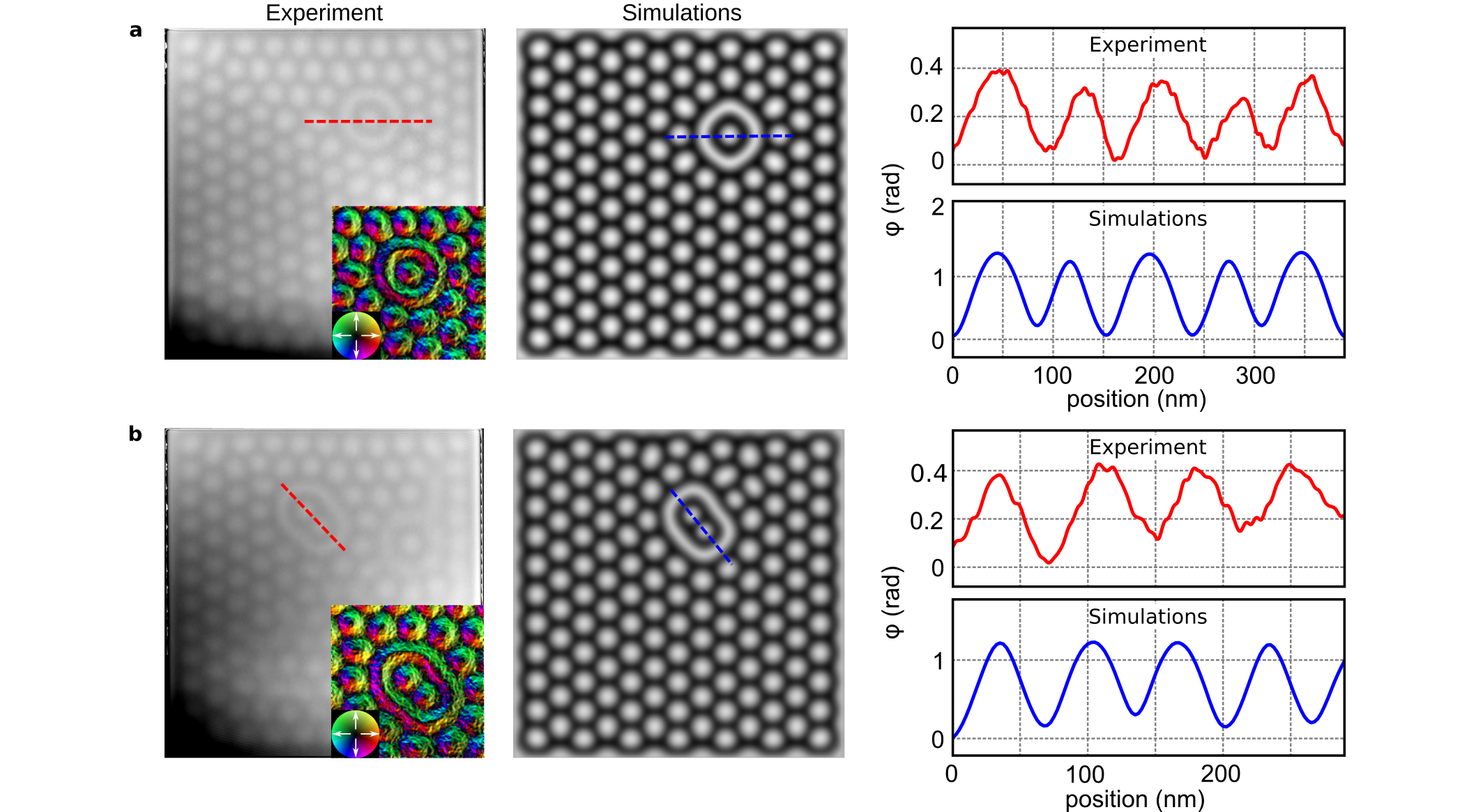}
    \caption{\small
    \textbf{Comparison between experimental and theoretical electron phase shift images of skyrmion bags in a 70-nm-thick FeGe plate.}
    \textbf{a}-\textbf{b} show experimental (left) and theoretical (middle) electron phase shift images for skyrmion bags with $Q~=~-1$ and $-2$, respectively, embedded in skyrmion lattices.
    Corresponding experimental Lorentz TEM images for the two skyrmion bags are shown in Fig.~\ref{Fig_collection}\textbf{b}-\textbf{c} and Supplementary Fig.~\ref{FigS-underfocus}\textbf{b}-\textbf{c}, respectively. 
    The insets to the experimental images show corresponding magnetic induction maps of each skyrmion bag and its surroundings. 
    The right column shows line profiles of the experimental (red) and theoretical (blue) phase shift across the center of each skyrmion bag.
    In the line profiles, the skyrmions and the outer rings of the skyrmion bags have nearly identical contrast, indicating that the rings, which are composed of two nested closed 180-degree domain walls, penetrate the full sample thickness. 
    This behavior is opposite to that observed in images of hopfion rings in thicker films of chiral magnets, in which the contrast from the ring is much weaker than that of the individual skyrmions.
    The experimental images were recorded at $T~=~250$~K, whereas in the micromagnetic simulations the parameters were adapted for $T~=~95$~K.
    The saturation magnetization and the strength of the magnetic induction field are therefore different for the experimental and theoretical images, resulting in a difference between the absolute values of the phase shift.
    }
    \label{Fig_comparison}
\end{figure*}

\begin{figure*}[ht] 
    \centering
    \includegraphics[width=17.5cm]{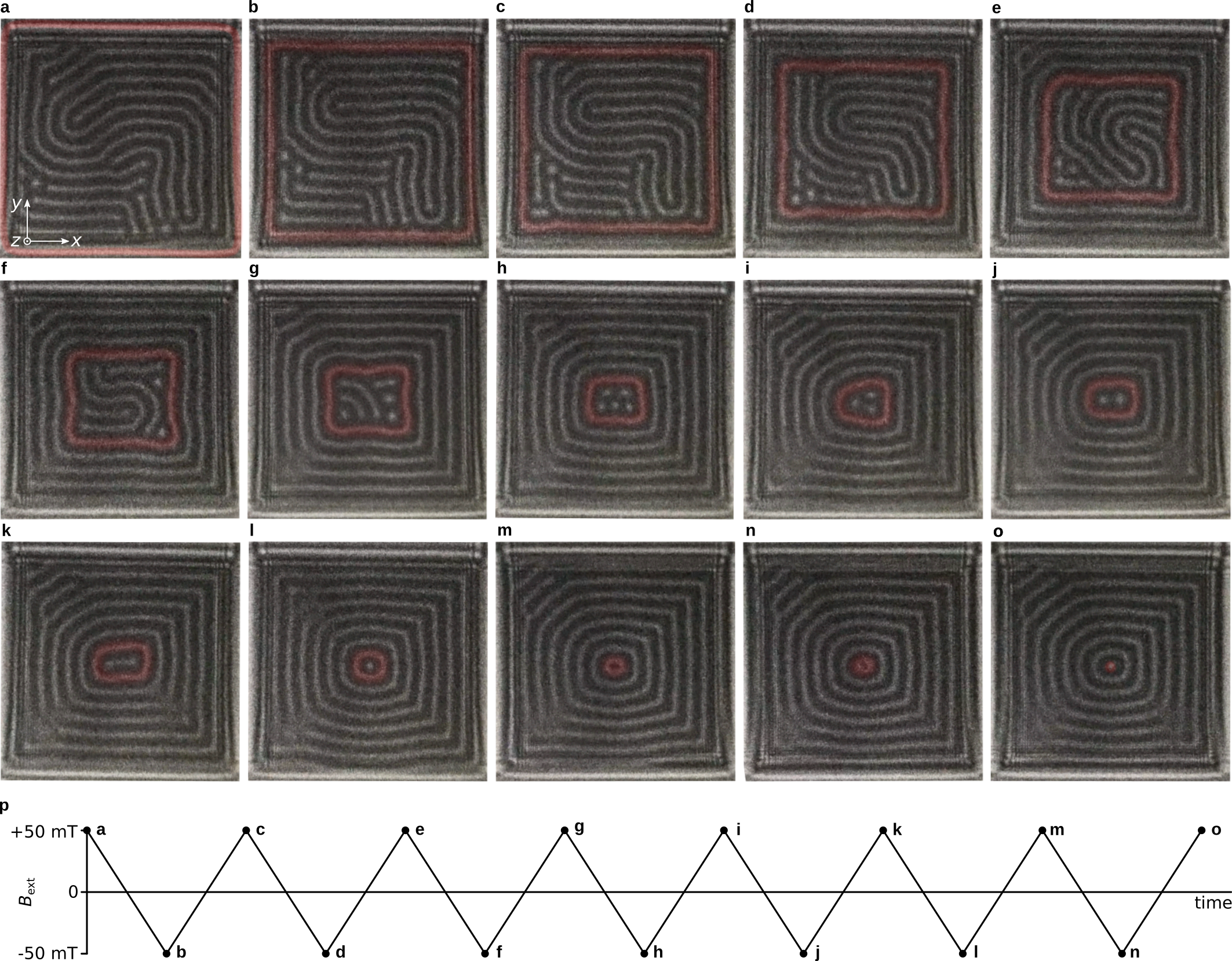}
    \caption{\small
    \textbf{Protocol for nucleation of magnetic skyrmion bags in a 180-nm-thick FeGe plate.}
    \textbf{a}-\textbf{o} show that, after several cycles of field swapping between +50 and -50~mT, helical edge modulations are generated and pushed towards the center of the sample (see the red line in each image).
    A closed loop of the helical spiral forms at the early stages of field swapping (see \textbf{b}), in part due to the perfection of the sample.
    In general, this state can be termed a ``skyrmion bag", albeit with helical spirals inside it.
    A skyrmion bag then forms with five skyrmions inside it (see \textbf{h}).
    Annihilation of the skyrmions in the bag is then observed during field swapping, with the number of skyrmions decreasing gradually to zero.
    The closed loop of the helical spiral (red line) turns into a skyrmion, leading to the formation of a skyrmion bag with one skyrmion inside it (see \textbf{o}). 
    The outer closed domain walls are also part of the bag in most images.     
    \textbf{p} shows a schematic illustration of the field swapping process.
    The frames show over-focus Lorentz TEM images recorded at $T$~=~250~K. The defocus distance is 400~$\mu$m.
    }
    \label{FigS-t180-nucleation}
\end{figure*}

\begin{figure*}[ht] 
    \centering
    \includegraphics[width=17.5cm]{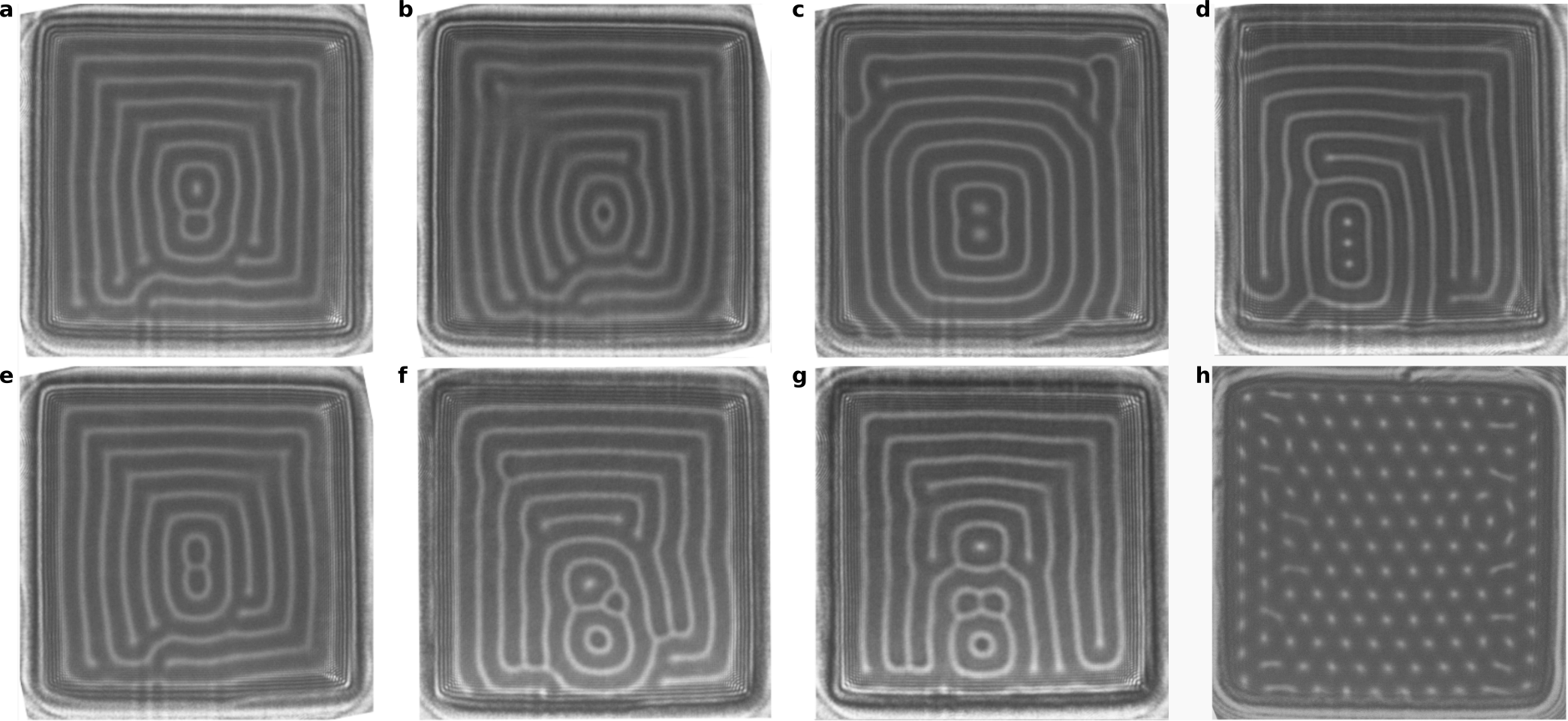}
    \caption{\small
    \textbf{Skyrmion bags embedded in the helical phase in a 180-nm-thick plate.}
    The frames show over-focus Lorentz TEM images recorded at a defocus distance of 400~$\mu$m. 
    For the skyrmion bags , the topological charge in \textbf{a} and \textbf{b}, $Q~=~0$, in \textbf{c}, $Q~=~-2$, in \textbf{d}, $Q~=~-3$, in \textbf{e} and \textbf{f}, $Q~=~2$ and in \textbf{g}, $Q~=~3$.
    The images in \textbf{a}-\textbf{g} were recorded in approximately zero external magnetic field.
    In a thick film studied at a high temperature ($T~>~T_\mathrm{a}$), all of the configurations shown in \textbf{a}-\textbf{g} undergo a transition to a skyrmion lattice phase in an increasing magnetic field.
    \textbf{h} shows a representative image of the skyrmion lattice phase recorded at $T~=~250$~K in an external magnetic field of 112~mT. 
    }
    \label{FigS-t180-bag}
\end{figure*}

\end{document}